\documentclass[aps,prd,preprint,preprintnumbers,amsmath,amssymb,nofootinbib]{revtex4-1}
\usepackage{epsf}
\usepackage{graphicx}
\usepackage{dcolumn}
\usepackage{bm}
\usepackage{amsfonts}
\usepackage[capitalize]{cleveref}
\usepackage{color}

\usepackage{centernot}

\begin{document}
%

\title{Spin-flavor oscillations of ultrahigh-energy cosmic neutrinos in interstellar space: The role of neutrino magnetic moments}

\author{Podist Kurashvili}
\affiliation{National Centre for Nuclear Research, Warsaw 00-681, Poland}
\email{pkurash@cern.ch}
\author{Konstantin A. Kouzakov}
\affiliation{Department of Nuclear Physics and Quantum
Theory of Collisions, Faculty of Physics, Lomonosov Moscow State University, Moscow 119991, Russia}%
\email{kouzakov@gmail.com}
\author{Levan Chotorlishvili}
\affiliation{Institut fur Physik, Martin-Luther-Universitat Halle-Wittenberg,
D-06099 Halle, Germany}
%
\author{Alexander I. Studenikin}
\affiliation{Department of Theoretical Physics, Faculty of
Physics, Lomonosov Moscow State University, Moscow 119991, Russia}%
\affiliation{Joint Institute for Nuclear Research, Dubna 141980, Moscow Region, Russia}%
\email{studenik@srd.sinp.msu.ru}


%
\begin{abstract}
A theoretical analysis of possible influence of neutrino magnetic moments on the propagation of ultrahigh-energy cosmic neutrinos in the interstellar space is carried out under the assumption of two-neutrino mixing. The exact solution of the effective equation for neutrino evolution in the presence of a magnetic field and matter is obtained, which accounts for four neutrino species corresponding to two different flavor states with positive and negative helicities. Using most stringent astrophysical bounds on the putative neutrino magnetic moment, probabilities of neutrino flavor and spin oscillations are calculated on the basis of the obtained exact solution. Specific patterns of spin-flavor oscillations are determined for neutrino-energy values characteristic of, respectively, the cosmogenic neutrinos, the Greisen-Zatsepin-Kuz'min (GZK) cutoff, and well above the cutoff.
\end{abstract}

\maketitle

%
\section{Introduction}
\label{sec:Introduction}
A remarkable progress of neutrino physics in the last decades has led to the discovery of neutrino oscillations, thereby showing that neutrinos are massive and mixed~\cite{Nobel2015}. As is well known, the neutrino massiveness supports the
assumption that neutrinos have nonzero electromagnetic characteristics~\cite{fujikawa80}. The studies of such nontrivial neutrino properties have a long history (see the recent review~\cite{bib:Giunti} and references therein). Owing to these studies many phenomena which can be induced by neutrino electromagnetic interactions have been predicted and described theoretically. In particular, such interactions are expected to generate observable effects in astrophysical environments~\cite{bib:Giunti,andp16}, where
neutrinos propagate over long distances in magnetic fields in vacuum and in matter. Among these effects the neutrino spin-flavor oscillations play a special role, for they feature the nonzero neutrino masses and electromagnetic properties on the one hand and the neutrino mixing on the other hand. 

In this scope it is worth mentioning that the neutrino spin oscillations $\nu^{L}\Leftrightarrow \nu^{R}$ induced by the neutrino magnetic moment interaction with the transversal magnetic field ${\bf B}_{\perp}$ was first considered in Ref.~\cite{Cisneros:1970nq}. Later, the spin-flavor oscillations $\nu^{L}_{e}\Leftrightarrow \nu^{R}_{\mu}$ in the presence of ${\bf B}_{\perp}$ in vacuum were discussed in Ref.~\cite{Schechter:1981hw}, and the importance of the matter effect was emphasized in Ref.~\cite{Okun:1986hi}. The effect of the resonant amplification of neutrino spin oscillations in the presence of ${\bf B}_{\perp}$ and matter was proposed in Refs.~\cite{Akhmedov:1988uk,Lim:1987tk}, and the impact of the longitudinal magnetic field ${\bf B}_{||}$ was discussed in Ref.~\cite{Akhmedov:1988hd}. The neutrino spin oscillations in the presence of a twisting magnetic field
were studied in Refs.~\cite{Vidal:1990fr, Smirnov:1991ia, Akhmedov:1993sh,Likhachev:1990ki,Dvornikov:2007aj,Dmitriev:2015ega}. 
Recently, a new approach to the description of neutrino spin and spin-flavor oscillations in the presence of an arbitrary constant magnetic field has been developed~\cite{Dmitriev:2015ega,Studenikin:2016zdx}. Within this approach the exact quantum stationary states in a magnetic field are used for classification of neutrino spin states, rather than the neutrino helicity states that have been  employed for this purpose within the customary approach in many previous works. 
In Ref.~\cite{Egorov:1999ah}, neutrino spin oscillations were considered in the presence of an arbitrary constant electromagnetic field $F_{\mu \nu}$. The treatment of neutrino spin oscillations in the circularly and linearly polarized electromagnetic waves and the superposition of an electromagnetic wave and a constant magnetic field can be found in Refs.~\cite{Lobanov:2001ar,Dvornikov:2001ez,Dvornikov:2004en}.

One of the important developments in the field of neutrino astrophysics is a search for ultrahigh-energy (UHE) cosmic neutrinos (even above PeV--EeV energies). These neutrinos are believed to be produced by reactions of UHE cosmic rays composed of protons and nuclei and are expected to provide information about cosmic accelerators and the high-energy, distant universe. It is well documented that the UHE cosmic ray spectrum varies smoothly up to an energy $E\sim40$\,EeV and drops off steeply beyond this point. This behavior is consistent with the Greisen-Zatsepin-Kuz'min (GZK) cutoff of about $50$\,EeV~\cite{GZK66}, which is set by production of pions in scattering of UHE cosmic rays off microwave photons. Thus, the primary signature of the UHE cosmic rays above the cutoff would be the neutrinos they produce in their interaction with the cosmic microwave background.

The UHE cosmic neutrinos can be detected with neutrino telescopes, for example, such as the IceCube neutrino observatory, which started operations in 2010 and already has reported observations of PeV cosmic neutrinos (see Ref.~\cite{Halzen_2016} and references therein). One of the major advantages of exploring the UHE neutrinos as astrophysical messengers is supposed to be their ability, as opposed to the case of charged particles, of traveling in straight lines in magnetic fields in space. This feature allows one to point back their intensively energetic sources in the sky, including active galactic nuclei, supernovae and associated phenomena like $\gamma$-ray
bursts, and compact objects such as black holes and neutron stars. At the same time, even though neutrinos are generally believed to be electrically neutral particles\footnote{The most stringent bound on the neutrino millicharge, which follows from the neutrality of matter, is $|e_\nu|\lesssim3\times10^{-21}e$~\cite{Raffelt99}.} they can still have nonzero magnetic moments. This means that the propagation of the UHE cosmic neutrinos can be influenced by the presence of magnetic fields due to the effect of spin oscillations. In particular, this influence can be substantial in the interstellar space of our galaxy, where the strength of a magnetic field takes on values of the order of few $\mu$G~\cite{Beck09}. Therefore, for both the current and the future studies with neutrino telescopes it is timely to examine how spin-flavor oscillations in the interstellar magnetic field can change the propagation pattern of UHE neutrinos.

The paper is organized as follows: in Sec.~\ref{sec:EquationOfMotion} we formulate the effective equation for neutrino evolution in the presence of a magnetic field and matter. Numerical results for probabilities of flavor and spin oscillations in the interstellar space are presented and discussed in Sec.~\ref{res}. Section~\ref{concl} summarizes this work. Finally, the Appendix is devoted to the exact solution of the evolution equation.

\section{General formulation}
\label{sec:EquationOfMotion}
We limit ourselves to the case of two Dirac neutrino physical states, $\nu_1$ and $\nu_2$, with masses $m_1$ and $m_2$. For treating neutrino evolution in the presence of a uniform magnetic field ${\bf B}$ and homogeneous matter in the ultrarelativistic limit, we employ a four-component basis of the helicity states $\nu_{1,s=\pm1}$ and $\nu_{2,s\pm1}$. The Schr\"odinger-like evolution equation is then given by
\begin{equation}
\label{eq:SchrodingerEquation}
i \frac{d  }{d t}
\begin{pmatrix}
\nu_{1,s=1} \\
\nu_{1,s=-1} \\
\nu_{2,s=1} \\
\nu_{2,s=-1}
\end{pmatrix}
 = H_{\rm eff} \begin{pmatrix}
\nu_{1,s=1} \\
\nu_{1,s=-1} \\
\nu_{2,s=1} \\
\nu_{2,s=-1}
\end{pmatrix}.
\end{equation}
The effective Hamiltonian $H_{\rm eff}$ consists of the vacuum and interaction parts,
\begin{equation}
\label{eq:HamiltonianFull}
H_{\rm eff}=H_{\rm vac} + H_{\rm mat} + H_{B},
\end{equation}
with $H_{\rm mat}$ and $H_{B}$ corresponding to the neutrino interaction with matter and a magnetic field, respectively. In what follows, we transform to the flavor basis using the relations
\begin{equation}
\label{eq:MassToFlavor}
\nu_e^{R,L}=\nu_{1,s=\pm1}\cos\theta+\nu_{2,s=\pm1}\sin\theta, \qquad \nu_\mu^{R,L}=-\nu_{1,s=\pm1}\sin\theta+\nu_{2,s=\pm1}\cos\theta,
\end{equation}
where $\nu_e^{R,L}$ and $\nu_\mu^{R,L}$ are electron and muon neutrino chiral states. In Eq.~(\ref{eq:MassToFlavor}) it is taken into account that in the discussed ultrarelativistic limit the chiral and helicity components practically coincide. In the flavor representation, the vacuum Hamiltonian acquires the form
\begin{equation}
\label{eq:HamiltonianFlavorRepresentation}
H_{\rm vac}^f=\omega
\begin{pmatrix}
-\cos 2 \theta & 0 & \sin 2 \theta & 0
\\
0 & -\cos 2 \theta & 0 & \sin 2 \theta
\\
\sin 2 \theta & 0 & \cos 2\theta & 0
\\
0 & \sin 2 \theta & 0 & \cos 2\theta
\end{pmatrix},
\end{equation}
where
\begin{equation}
\label{eq:DeltaM}
\omega= \frac{\Delta m^2}{4 E_\nu}, \qquad \Delta m^2 =m_2^2 - m_1^2,
\end{equation}
with $E_\nu$ being the neutrino energy. The neutrino-matter interaction in the flavor basis~(\ref{eq:MassToFlavor}) is described by the Hamiltonian
\begin{equation}
\label{eq:InteractionHamiltonian}
H_{\rm mat}^f=
 \lambda
\begin{pmatrix}
0 & 0 & 0 & 0
\\
0 & 1 & 0 & 0
\\
0 & 0 & 0 & 0
\\
0 & 0 & 0 & -1
\end{pmatrix},
\end{equation}
where $\lambda= \frac {1}{\sqrt{2}}\, G_F n_e$, with the Fermi constant $G_F$ and the net electron density $n_e=n_{e^-}-n_{e^+}$. The Hamiltonian of the neutrino interaction with a magnetic field in the flavor representation can be presented as~\cite{bib:Fabbricatore}:
\begin{equation}
\label{eq:H_EM}
H_{B}^f=
\begin{pmatrix}
\displaystyle -\left(\frac{\mu}{\gamma}\right)_{ee} {B_\parallel} &&
\mu_{ee}B_{\perp} &&
\displaystyle -\left(\frac{\mu}{\gamma}\right)_{ e\mu}{B_\parallel} &&
\mu_{e\mu}B_\perp
\\
\mu_{ee}B_\perp &&
\displaystyle \left(\frac{\mu}{\gamma}\right)_{ee}{B_\parallel} &&
\mu_{e\mu} B_\perp &&
\displaystyle \left(\frac{\mu}{\gamma}\right)_{e\mu}{B_\parallel}
\\
\displaystyle -\left(\frac{\mu}{\gamma}\right)_{e\mu}{B_\parallel} &&
\mu_{e\mu}B_{\perp} &&
\displaystyle -\left(\frac{\mu}{\gamma}\right)_{\mu \mu}{B_\parallel}
&&
\mu_{\mu \mu} B_\perp
\\
\mu_{e\mu}B_{\perp} &&
\displaystyle \left(\frac{\mu}{\gamma}\right)_{e\mu}{B_\parallel} &&
\mu_{\mu \mu}B_\perp &&
\displaystyle \left(\frac{\mu}{\gamma}\right)_{\mu\mu}{B_\parallel}
\end{pmatrix},
\end{equation}
where $B_\parallel$ and $B_\perp$ are the parallel and transverse magnetic-field components with respect to the neutrino velocity, and the magnetic moments $\tilde{\mu}_{\ell\ell'}$ and $\mu_{\ell\ell'}$ ($\ell,\ell'=e,\mu$) are related to those in the mass representation $\mu_{jk}$ ($j,k=1,2$) as follows:
\begin{align}
\label{eq:MuPrime}
\mu_{ee}&=\mu_{11} \cos^2 \theta +\mu_{22} \sin^2 \theta
+\mu_{12} \sin 2\theta,
\nonumber
\\
\mu_{e\mu}&=\mu_{12}\cos 2\theta + \frac{1}{2}
\left( \mu_{22} - \mu_{11}\right)\sin 2\theta,
\\
\mu_{\mu\mu}&=\mu_{11} \sin^2 \theta
+\mu_{22} \cos^2 \theta-\mu_{12} \sin 2\theta,
\nonumber
\end{align}
and
\begin{align}
\label{eq:MuTilde}
\left(\frac{\mu}{\gamma}\right)_{ee} &=
\frac{\mu_{11}}{\gamma_{1}}\,\cos^2 \theta +
\frac{\mu_{22}}{\gamma_{2}}\,\sin^2 \theta +
\frac{\mu_{12}}{\gamma_{12}}\,\sin 2\theta,
\nonumber
\\
\left(\frac{\mu}{\gamma}\right)_{e\mu} &=
\frac{\mu_{12}}{\gamma_{12}}\,\cos 2\theta
+\frac{1}{2}\left(
\frac{\mu_{22}}{\gamma_{2}}-\frac{\mu_{11}}{\gamma_{1}}
\right)\sin 2\theta,
\\
\left(\frac{\mu}{\gamma}\right)_{\mu\mu}&=
\frac{\mu_{11}}{\gamma_{1}}\,\sin^2 \theta
+\frac{\mu_{22}}{\gamma_{2}}\,\cos^2 \theta
-\frac{\mu_{12}}{\gamma_{12}}\,\sin 2\theta.
\nonumber
\end{align}
Here $\gamma_1$ and $\gamma_2$ are the Lorenz factors of the massive neutrinos, and
\begin{equation}
\label{eq:GammaDefinition}
\frac{1}{\gamma_{12}}=\frac{1}{2}\left(\frac{1}{\gamma_1}+\frac{1}{\gamma_2}\right).
\end{equation}

Let us first briefly recapitulate the main results for the case of the absence of a magnetic field (or, more generally, when the neutrino interaction with a magnetic field is absent). In such a situation, the neutrino states with different chiralities decouple, so that for the left-chiral states in the flavor basis the evolution equation acquires the form
\begin{equation}
\label{eq:SystemeZeroB}
 i\frac{d}{dt}
\begin{pmatrix}
\nu_e^L \\ \nu_\mu^L
\end{pmatrix}=
\begin{pmatrix} -\omega \cos 2 \theta + \lambda & \omega \sin 2\theta \\
\omega \sin 2\theta & \omega \cos 2 \theta - \lambda
\end{pmatrix}\begin{pmatrix}
\nu_e^L \\ \nu_\mu^L
\end{pmatrix}.
\end{equation}
The eigenvalues of (\ref{eq:SystemeZeroB}) are $\omega_{1,2}=\pm\omega_m$, where
\begin{equation}
\label{eq:EigenvaluesZeroB}
\omega_m=\sqrt{\left(
\omega \cos 2 \theta -\lambda
\right)^2 + \omega^2 \sin^2 2 \theta}.
\end{equation}
If initially the neutrino is in the $\nu_e^L$ state, the flavor-change probability is given by~\cite{bib:Wolfenstein}
\begin{equation}
\label{Eq:Pflavor_matter}
P_{\nu_e^L\to\nu_\mu^L}(t)=\sin^2{2\theta_m} \sin^2{\omega_m t},
\end{equation}
with
\begin{equation}
\label{eq:ThetaM}
\sin 2\theta_m = \frac{\omega}{\omega_m}\,\sin{2\theta}.
\end{equation}

When the neutrino interacts both with matter and with a magnetic field, the evolution equation~(\ref{eq:SchrodingerEquation}) in the flavor basis leads to the following homogeneous system of first-order linear differential equations:
\begin{align}
\label{eq:SystemNonZeroB}
\begin{cases}
\displaystyle i\frac{d\nu_{e}^R}{d t}=
\left[
-\omega \cos 2\theta
- \left(\frac{\mu}{\gamma}\right)_{ee}
B_\parallel\right]\nu_{e}^R
+
\mu_{ee}B_{\perp} \nu_{e}^L
+\left[\omega \sin 2\theta
-\left(\frac{\mu}{\gamma}\right)_{e\mu}B_\parallel
 \right]  \nu_{\mu}^R
+
\mu_{e\mu}B_\perp \nu_{\mu}^L,
\\[3mm]
\displaystyle i \frac{d\nu_{e}^L}{d t} =
\mu_{ee}B_\perp \nu_{e}^R+
\left[
- \omega_m \cos 2\theta_m +
\left(\frac{\mu}{\gamma}\right)_{ee}B_\parallel \right] \nu_{e}^L
+ \mu_{e\mu} B_\perp \nu_{\mu}^R+
\left[ \omega \sin 2\theta +
\left(\frac{\mu}{\gamma}\right)_{e\mu}B_\parallel
 \right]\nu_{\mu}^L,
\\[3mm]
\displaystyle i \frac{d\nu_{\mu}^R}{d t}  =
\left[\omega \sin 2\theta
-\left(\frac{\mu}{\gamma}\right)_{e\mu}B_\parallel
\right] \nu_{e}^R
+\mu_{e\mu}B_{\perp} \nu_{e}^L
+ \left[\omega \cos 2\theta
-\left(\frac{\mu}{\gamma}\right)_{\mu \mu}B_\parallel
\right] \nu_{\mu}^R +
\mu_{\mu \mu} B_\perp \nu_{\mu}^L,
\\[3mm]
\displaystyle i\frac{d\nu_{\mu}^L}{\partial t} =
\mu_{e\mu}B_{\perp} \nu_{e}^R +
\left[ \omega\sin 2\theta+  \left(\frac{\mu}{\gamma}\right)_{e\mu}B_\parallel
 \right] \nu_{e}^L + \mu_{\mu \mu}B_\perp \nu_{\mu}^R +
\left[\omega_m \cos 2\theta_m  + \left(\frac{\mu}{\gamma}\right)_{\mu \mu}B_\parallel
\right] \nu_{\mu}^L.
\end{cases}
\end{align}
The above system is equivalent to a fourth-order homogeneous linear differential equation. This means that the general solution contains four linearly independent parts, namely
\begin{equation}
\label{GeneralSolution}
\mathbb{\nu}^f(t)=\sum_{\alpha=1}^4C_\alpha\nu_\alpha^fe^{-i\tilde{\omega}_\alpha t}, 
\end{equation}
where $\nu^f\equiv(\nu_{e}^{R},\nu_{e}^{L},\nu_{\mu}^{R},\nu_{\mu}^{L})^T$, $\tilde{\omega}_{\alpha=1,2,3,4}$ and $\nu_{\alpha=1,2,3,4}^f$ are
the eigenvalues and eigenstates of the effective Hamiltonian $H^f_{eff}$ (see the Appendix), and
$$
 C_\alpha=\langle\nu_\alpha^f|\mathbb{\nu}^f(0)\rangle, \qquad \sum_{\alpha=1}^4|C_\alpha|^2=1.
$$
\section{Results and discussion}
\label{res}
In this section we present and discuss numerical results for oscillations of UHE cosmic neutrinos in the interstellar space. We neglect the neutrino interaction with the longitudinal magnetic-field component, setting $(\mu/\gamma)B_\parallel=0$. The latter is justified by large $\gamma$ values for UHE neutrinos. The magnetic field strength is set to the value $B=2.93\,\mu$G that has recently been measured by NASA's Interstellar Boundary Explorer~\cite{NASA_ISMF}. The neutrino coupling to a magnetic field depends on the values of the diagonal and transition magnetic moments. In the present calculations, the following case of neutrino magnetic-moment values is inspected: $\mu_{11}=\mu_{22}=\mu_{12}=\mu_\nu$, yielding $\mu_{ee}=\mu_\nu(1+\sin2\theta)$, $\mu_{\mu\mu}=\mu_\nu(1-\sin2\theta)$, and $\mu_{e\mu}=\mu_\nu\cos2\theta$. For the putative magnetic moment we use the values $\mu_\nu=2.6\times10^{-12}\mu_B$ and $\mu_\nu=4.5\times10^{-12}\mu_B$, which correspond to the upper bounds (at 67\% CL and 95\% CL, respectively) on the neutrino dipole magnetic moments obtained from the constraints on the possible delay of helium ignition of a red giant star in globular clusters due to the cooling induced by the plasmon-decay energy loss~\cite{Viaux_2013}. The neutrino interaction with a magnetic field is then of the order of $\mu_\nu B \sim 10^{-26}-10^{-25}$\,eV. Since in the interstellar space $n_e\lesssim10^{6}$\,cm$^{-3}$, the matter effects can safely be ignored in the calculations, i.e., we put $\lambda=0$ and, accordingly, $\omega_m=\omega$.  The square mass difference is taken from the solar neutrino measurements, $\Delta m^2 =\Delta m_{sol}^2=7.37\times 10^{-5}$\,eV$^2$, and the vacuum mixing angle is $\sin^2 \theta = 0.297$~\cite{PDG2016}. All numerical calculations are performed for the case when the initial neutrino state is $\nu^f(0)=\nu_e^L$.

\begin{figure}
\centering
\includegraphics[width=0.8\textwidth]{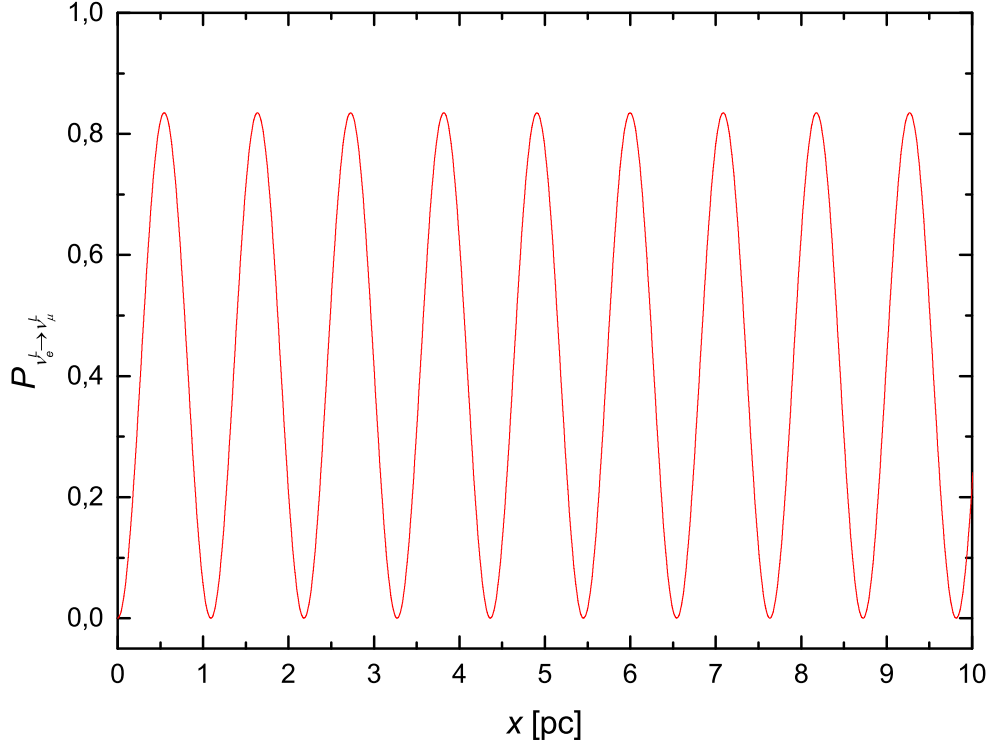}
\caption{\label{fig1} The neutrino flavor-change probability as a function of the distance $x$ traveled by an $1$-EeV neutrino in vacuum.}
\end{figure}

Fig.~\ref{fig1} shows the flavor-change probability $P_{\nu_e^L\to\nu_\mu^L}$ for the neutrino propagating in vacuum with an energy typically anticipated for cosmogenic neutrinos, $E_\nu=1$\,EeV~\cite{Meures_book,bib:Ishihara}. According to Eq.~(\ref{Eq:Pflavor_matter}), the dependence of $P_{\nu_e^L\to\nu_\mu^L}$ on the neutrino propagation distance $x$ in this case is determined by the following formula:
\begin{equation}
\label{Eq:Pflavor_vac} P_{\nu_e^L\to\nu_\mu^L}(x)=\sin^22\theta\,\sin^2\left(\frac{\pi x}{L_{\rm vac}}\right),
\end{equation}
where the vacuum oscillation length is $L_{\rm vac}=4\pi E_\nu/\Delta m^2=1.09$\,pc. The latter value appears to be by orders of magnitude smaller than the Sun's distance from the Galactic Center ($\approx8$\,kpc). The neutrino flavor-change probability~(\ref{Eq:Pflavor_vac}) reaches its maximal value $P_{\nu_e^L\to\nu_\mu^L}^{\,\rm max}=\sin^22\theta=0.835$ at distances $x_k=(2k+1)L_{\rm vac}/2$, where $k\in\mathbb{N}_0$.

\begin{figure}
\centering
\includegraphics[width=0.8\textwidth]{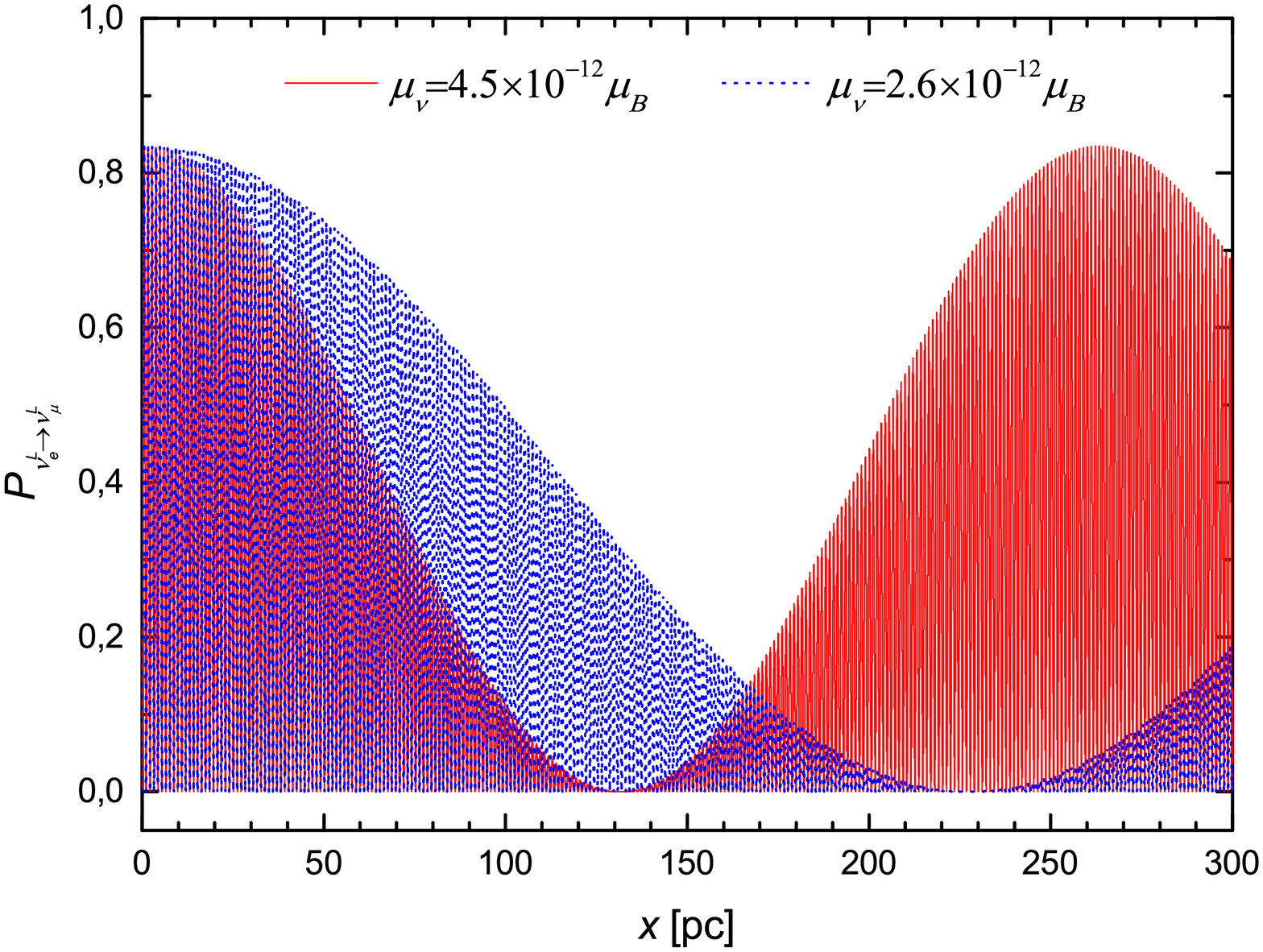}\\[.5cm]
\includegraphics[width=0.8\textwidth]{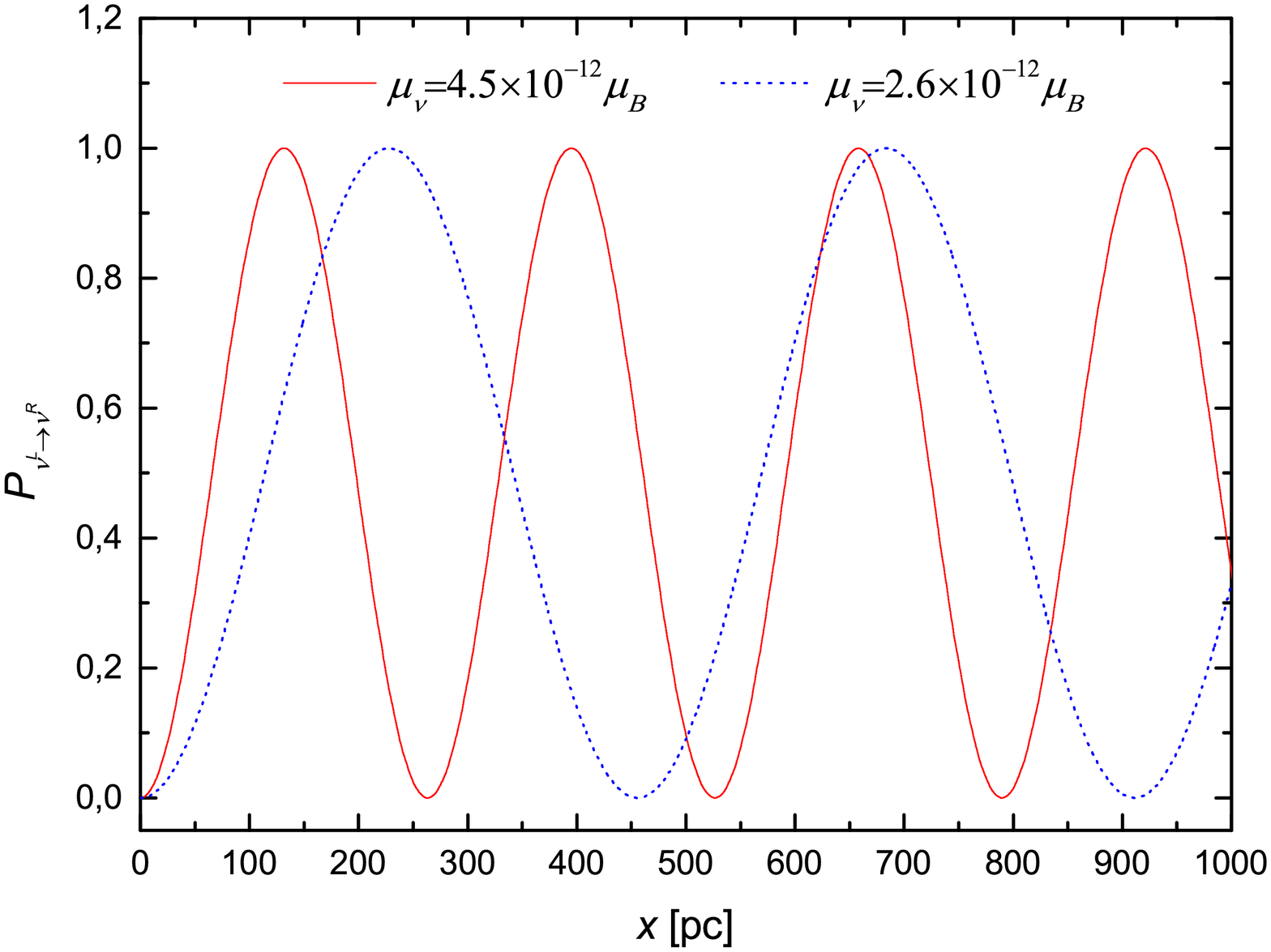}
\caption{\label{fig2} The neutrino flavor-change (top panel) and spin-flip (bottom panel) probabilities as functions of the distance $x$ traveled by an $1$-EeV neutrino interacting with an interstellar magnetic field.
}
\end{figure}

If the propagating neutrino interacts with an interstellar magnetic field the neutrino spin oscillations become possible, which can influence the flavor-change probability~(\ref{Eq:Pflavor_vac}) due to the $\nu^L\to\nu^R$ conversion process. The results for $P_{\nu_e^L\to\nu_\mu^L}$ in this case are presented in Fig.~\ref{fig2}. They exhibit fast oscillations with the period $L_{\rm vac}$ which are modulated by a slowly changing envelope curve that depends on the $\mu_\nu$ value. This behavior can be explained by the simple formula
\begin{equation}
\label{Eq:Pflavor_B} P_{\nu_e^L\to\nu_\mu^L}(x)=\left[1-P_{\nu^L\to\nu^R}(x)\right]\sin^22\theta\,\sin^2\left(\frac{\pi x}{L_{\rm vac}}\right).
\end{equation}
The envelope curve is determined by the neutrino spin-flip probability $P_{\nu^L\to\nu^R}=P_{\nu^L_e\to\nu^R_e}+P_{\nu^L_e\to\nu^R_\mu}$ which is also shown in Fig.~\ref{fig2}. The $P_{\nu^L\to\nu^R}$ results are very well fitted by
\begin{equation}
\label{Eq:Pspin_B} P_{\nu^L\to\nu^R}(x)=\sin^2\left(\frac{\pi x}{L_{B}}\right),
\end{equation}
where $L_B=\pi/\mu_\nu B$ is the magnetic oscillation length that takes on values of 263.2\,pc and 455.6\,pc for $\mu_\nu=4.5\times10^{-12}\mu_B$ and $\mu_\nu=2.6\times10^{-12}\mu_B$, respectively.



\begin{figure}
\centering
\includegraphics[width=0.8\textwidth]{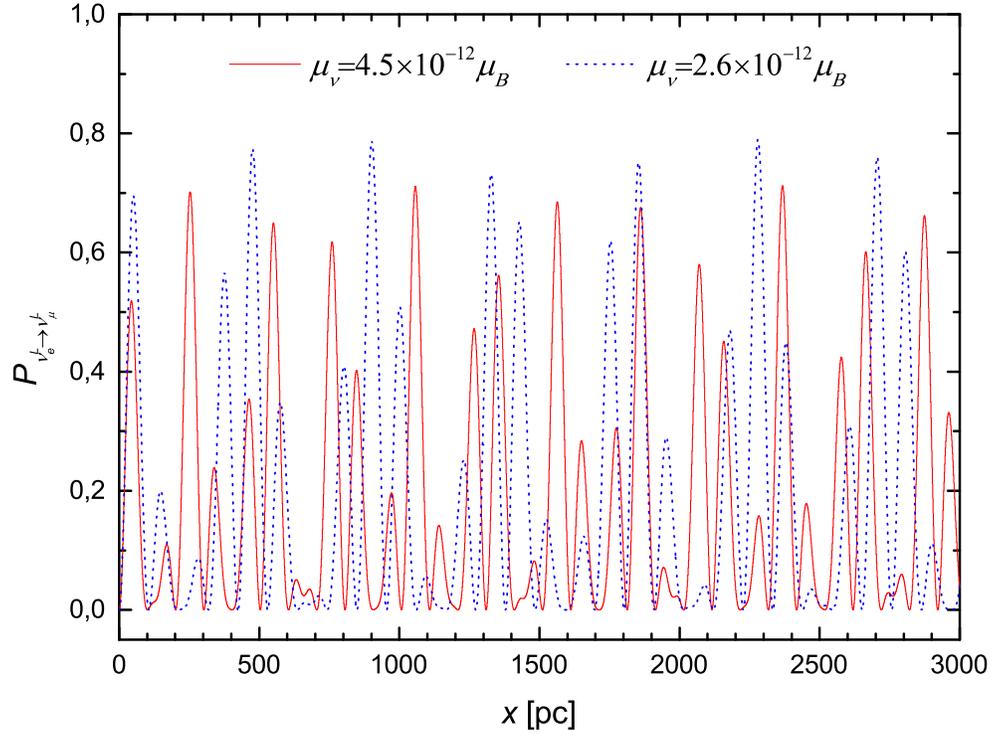}\\[.5cm]
\includegraphics[width=0.8\textwidth]{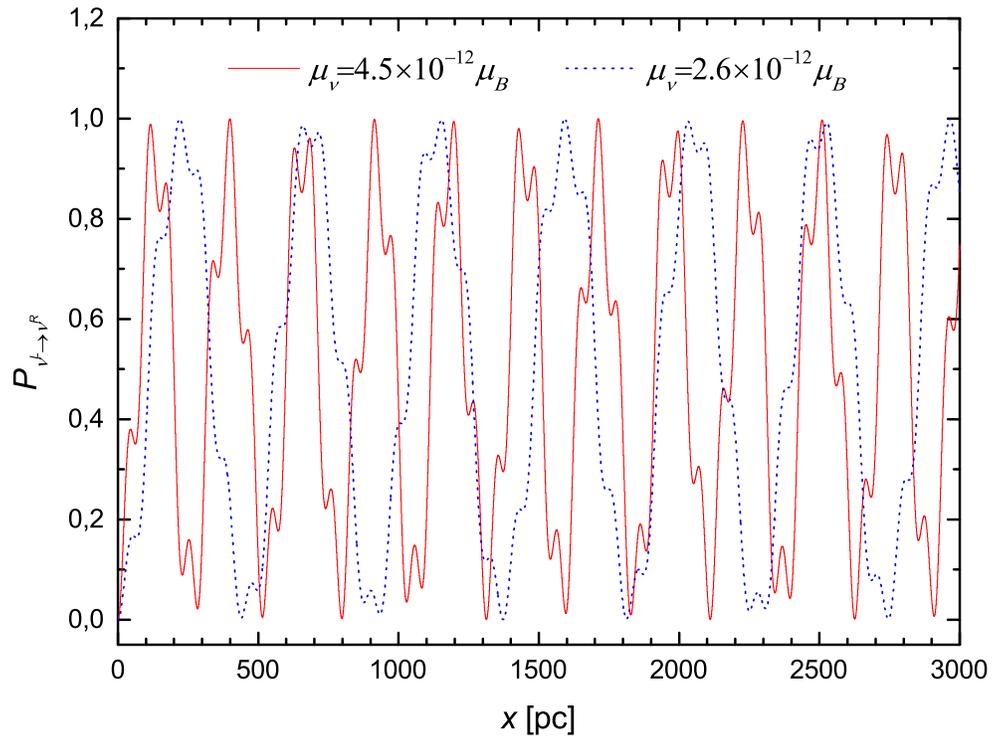}
\caption{\label{fig5} The same as in Fig.~\ref{fig2}, but when $E_\nu=100$\,EeV.}
\end{figure}

The simple behaviors of the neutrino flavor-change and spin-flip probabilities, (\ref{Eq:Pflavor_B}) and (\ref{Eq:Pspin_B}), owe to the fact that $L_{B}\gg L_{\rm vac}$. Indeed, in such a situation the spin oscillations take place in the adiabatic regime compared with the flavor oscillations. It allows one to draw an analogy with the amplitude modulation of a sinusoidal wave signal, where the wave with the vacuum oscillation frequency $\omega=\pi/L_{\rm vac}$ plays a role of the carrier wave and that with a much lower frequency, $\omega_B=\pi/L_{B}$, represents the modulation waveform. This picture becomes inapplicable when the oscillation lengths $L_B$ and $L_{\rm vac}$ are comparable. For example, this can be the case for neutrino energies of the order of the GZK cutoff, such as $E_\nu=100$\,EeV. The corresponding value of the vacuum oscillation length is $L_{\rm vac}=109$\,pc, 
which is of the same order of magnitude as the $L_B$ values under consideration. It can be seen from Fig.~\ref{fig5} that the oscillation pattern of the flavor-change probability for a 100-EeV neutrino interacting with an interstellar magnetic field qualitatively differs from that in Fig.~\ref{fig2}. In particular, it exhibits peaks of rather irregular intensity, whose maximums not only do not reach the value $P_{\nu_e^L\to\nu_\mu^L}^{\,\rm max}=0.835$, in contrast to the $E_\nu=1$\,EeV case, but even do not exceed the value of 0.8. This reflects an interplay between the flavor-change and spin-flip processes, which is also manifested in the $E_\nu=100$\,EeV results for the $P_{\nu^L\to\nu^R}$ probability presented in Fig.~\ref{fig5}. It can be clearly seen that the $x$ dependence of this probability does not follow Eq.~(\ref{Eq:Pspin_B}) and has a rather complicatedly-shaped waveform, which is only approximately periodic.

%

\begin{figure}
\centering
\includegraphics[width=0.8\textwidth]{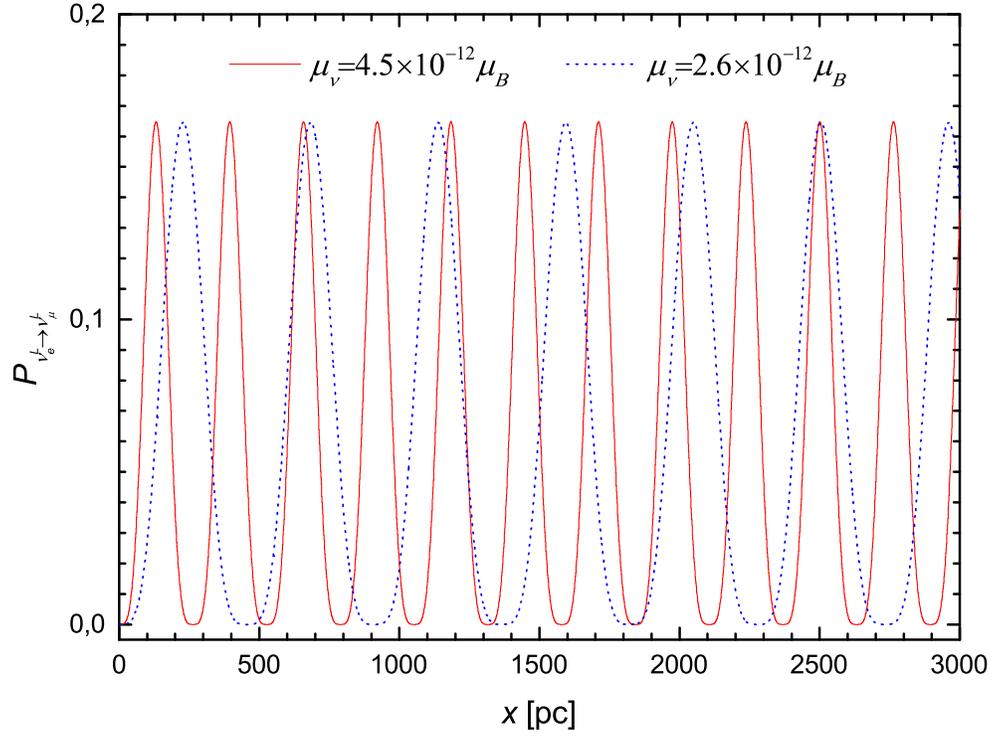}\\[.5cm]
\includegraphics[width=0.8\textwidth]{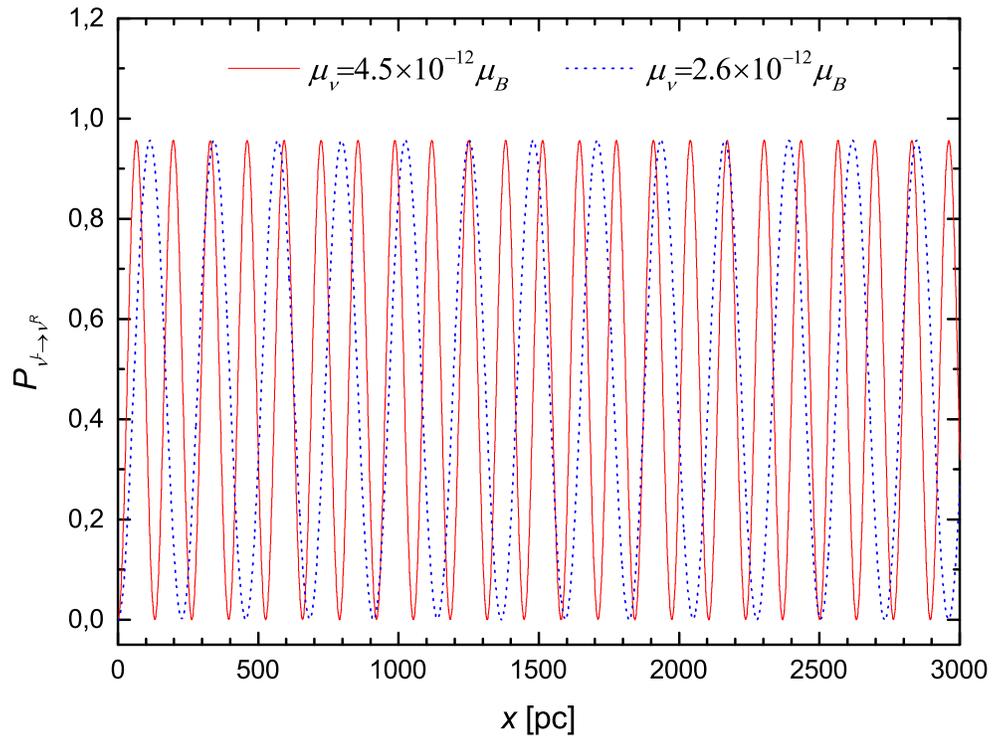}
\caption{\label{fig7} The same as in Fig.~\ref{fig2}, but when $E_\nu=10$\,ZeV.}
\end{figure}

So far we have discussed the $L_B\gg L_{\rm vac}$ and $L_B\sim L_{\rm vac}$ cases. However, if the neutrino energy is well above the GZK cutoff, the $L_{\rm vac}$ scale can be substantially larger than $L_B$. Fig.~\ref{fig7} shows the neutrino flavor-change and spin-flip probabilities when $E_\nu=10$\,ZeV, so that the vacuum oscillation length is $L_{\rm vac}=10.9$\,kpc ($\gg L_B$), which even exceeds the Sun's distance from the Galactic Center. The results are qualitatively different from those both in Fig.~\ref{fig2} and in Fig.~\ref{fig5}. Namely, the flavor-change probability oscillates with a period of $L_B$ and its maximal value is substantially suppressed, and the spin-flip probability oscillates with a period of $L_B/2$ and its maximal value is about 4--5\% less than unity. These findings can be explained by solving the system of first-order linear differential equations~(\ref{eq:SystemNonZeroB}) in the limit $\omega_m=\omega=0$ (this amounts to $L_{\rm vac}\to\infty$). One obtains the following eigenvalues: $\tilde\omega_{1,2}=0$ and $\tilde\omega_{3,4}=\pm2\mu B$. The corresponding neutrino flavor-change and spin-flip probabilities are given by the following expressions:
\begin{equation}
\label{Eq:w=0}
P_{\nu_e^L\to\nu_\mu^L}(x)=\cos^22\theta\,\sin^2\left(\frac{\pi x}{L_{B}}\right), \qquad
P_{\nu^L\to\nu^R}(x)=\frac{1}{2}\,(1+\sin2\theta)\sin^2\left(\frac{2\pi x}{L_{B}}\right),
\end{equation}
with $\cos^22\theta=0.165$ and $(1+\sin2\theta)/2=0.957$. The strong suppression of the $\nu_e^L\to\nu_\mu^L$ conversion probability in the discussed limit stems from the fact that it takes place solely through the spin-flip processes $\nu_e^L\to\nu_\mu^R\to\nu_\mu^L$ and $\nu_e^L\to\nu_e^R\to\nu_\mu^L$. The appearance of $\sin2\theta$ and $\cos2\theta$ in Eq.~(\ref{Eq:w=0}) is due to the neutrino mixing which is reflected in the relations between the neutrino magnetic moments in the flavor and mass bases~(\ref{eq:MuPrime}).

\section{Summary and conclusions}
\label{concl}
We have performed a theoretical analysis of possible effects of neutrino magnetic moments on the propagation of UHE cosmic neutrinos in the interstellar space. An effective equation for neutrino evolution in the presence of a constant magnetic field and homogeneous matter has been formulated within the framework of two Dirac neutrino states. Using the exact solution of this equation, we have calculated probabilities of UHE neutrino flavor and spin oscillations for several neutrino-energy values, namely $E_\nu=1$\,EeV, 100\,EeV and 10\,ZeV. Characteristic behaviors of these probabilities depending on the relative scale of the vacuum and magnetic oscillation lengths, $L_{\rm vac}$ and $L_B$, have been determined. It has been found that when $L_{\rm vac}\ll L_B$ the probability of flavor oscillations exhibits a vacuumlike dependence on the neutrino travel distance modulated by the spin oscillations with a period of $L_B$. If $L_{\rm vac}\gg L_B$, the probability of flavor oscillations has been found to be strongly suppressed and having a period of $L_B$ that turns out to be twice larger than that of spin oscillations in this case. Finally, in the $L_{\rm vac}\sim L_B$ regime one deals with a complex interplay between the flavor and spin oscillation processes. In all the inspected regimes both the flavor and spin oscillation probabilities have been found to be very sensitive to the value of the neutrino magnetic moment $\mu_\nu$.

Some comments should be made about the case of Majorana neutrinos. The major difference, as compared to the Dirac neutrinos studied in the present work, consists not only in the interaction of the right-handed Majorana neutrinos with matter but also in general properties of their magnetic moments $\mu^M_{ij}$. The matrix of the magnetic moments for Majorana neutrinos is antisymmetric and Hermitian (see, for instance, Ref.~\cite{bib:Giunti} for details), so that in the discussed scenario of two-neutrino mixing one has $\mu^M_{11}=\mu^M_{22}=0$ and $\mu^M_{12}=-\mu^M_{21}=-(\mu^{M}_{12})^*$. From the latter relationship it follows that the transition magnetic moment $\mu^M_{12}$ is a purely imaginary quantity and, using the putative magnetic moment $\mu_\nu$, can be parametrized as $\mu^M_{12}=\pm i\mu_\nu$. It is straightforward to derive expressions for the flavor-change and spin-flip probabilities in the absence of the neutrino interaction with matter, which were examined in the previous section, in the Majorana-neutrino case. The resulting expressions are
$$
P^M_{\nu_e^L\to\nu_\mu^L}(x)=\left(\frac{\tilde{L}}{L_{\rm vac}}\right)^2\sin^22\theta\,\sin^2\left(\frac{\pi x}{\tilde{L}}\right), \qquad
P^M_{\nu^L\to\nu^R}(x)=\left(\frac{\tilde{L}}{L_{B}}\right)^2\sin^2\left(\frac{\pi x}{\tilde{L}}\right),
$$
where $\tilde{L}=L_{\rm vac}L_{B}/\sqrt{L_{\rm vac}^{2}+L_B^{2}}$. Both probabilities oscillate with the same period, in contrast to the Dirac case. Another distinct feature in comparison with the Dirac case is that in the limit $L_{B}/L_{\rm vac}\to0$ the probability of flavor oscillations vanishes [cf. Eq.~(\ref{Eq:w=0})]. In principle, one might use the indicated marked differences to distinguish between Dirac and Majorana neutrinos by measuring the flavor composition and/or deficit of active UHE neutrinos from similar astrophysical sources located at different distances from the Earth. A more direct way, which makes use of the effect of the spin conversion $\nu^L\to\nu^R$ ($\bar{\nu}^R\to\bar{\nu}^L$), is to measure antineutrinos (neutrinos) from the sources of neutrinos (antineutrinos).

The results of the present analysis can be important for searches of UHE cosmic neutrinos and interpretation of the data of neutrino telescopes. They also can be important for the search of neutrino magnetic moments and studies of neutrino spin-flavor oscillations' effects in astrophysics.

\begin{acknowledgments}
We are grateful to Alexey Lokhov, Alexander Grigoriev and Pavel Pustoshny for useful discussions and valuable comments. This work was supported by the Russian Foundation for Basic Research under Grants No. 16-02-01023\,A and No. 17-52-53133\,GFEN\_a.
\end{acknowledgments}
\appendix
\section{The eigenvalues and eigenstates of the effective Hamiltonian}
\label{app-A}
The eigenvalues of the effective Hamiltonian $H_{\rm eff}^f$ are determined by the roots of the characteristic equation:
\begin{equation}
\label{eq:CharacteristicEq}
\det\left(H_{\rm eff}^f-\tilde{\omega}I\right)=0.
\end{equation}
Since ${\rm Tr}\left(H_{\rm eff}^f\right)=0$, from~(\ref{eq:CharacteristicEq}) one obtains a quartic equation of the standard form,
\begin{equation}
\label{eq:QuarticEq}
\tilde{\omega}^4+p\tilde{\omega}^2+q\tilde{\omega}+r=0,
\end{equation}
with the coefficients~\cite{Zadeh_book}
\begin{eqnarray}
p=-\frac{1}{2}\,{\rm Tr}\left(H_{\rm eff}^{f\,2}\right),
\qquad q=-\frac{1}{3}\,{\rm Tr}\left(H_{\rm eff}^{f\,3}\right), \qquad r=\det\left(H_{\rm eff}^f\right). 
\end{eqnarray}
The roots of~(\ref{eq:QuarticEq}) are given by~\cite{Abramowitz_book,Beyer_book}
\begin{equation}
\label{eq:RootsOfQuarticEq}
\tilde{\omega}_{1,2}=\frac{1}{2}\,R\pm\sqrt{-R^2-2p-\frac{2q}{R}}, \qquad
\tilde{\omega}_{3,4}=-\frac{1}{2}\,R\pm\sqrt{-R^2-2p+\frac{2q}{R}},
\end{equation}
where
$$
R=\sqrt{y_1-p}\neq0,
$$
with $y_1$ being a real root of the resolvent cubic equation
$$
y^3-py^2-2qy+4pr-q^2=0.
$$
If $R=0$, then
\begin{equation}
\label{eq:RootsOfQuarticEqR=0}
\tilde{\omega}_{1,2}=\pm\sqrt{-2p+2\sqrt{y_1^2-4r}}, \qquad
\tilde{\omega}_{3,4}=\pm\sqrt{-2p-2\sqrt{y_1^2-4r}}.
\end{equation}

The eigenstate of $H_{\rm eff}^f$ that corresponds to the eigenvalue $\tilde{\omega}_\alpha$ can be presented as
\begin{equation}
\label{eq:Eigenstate}
\nu_\alpha^f=\frac{1}{\sqrt{N_j^\alpha}}
\begin{pmatrix}
A_{j1}^\alpha \\ A_{j2}^\alpha \\ A_{j3}^\alpha \\ A_{j4}^\alpha
\end{pmatrix}, \qquad N_j^\alpha=\left|A_{j1}^{\alpha}\right|^2 + \left|A_{j2}^{\alpha}\right|^2 + \left|A_{j3}^{\alpha}\right|^2 + \left|A_{j4}^{\alpha}\right|^2,
\end{equation}
where $A_{jk}^{\alpha}$ is the $(j,k)$ cofactor of the matrix $||H_{\rm eff}^f-\tilde{\omega}_\alpha I||$, whose rank is supposed to be $r=3$. Clearly, the choice of the $j$ value in Eq.~(\ref{eq:Eigenstate}) is restricted by the condition that at least one of the cofactors $A_{jk=1,2,3,4}^{\alpha}$ must be nonzero.

If the neutrino state at $t=0$ is
\begin{equation}
\label{eq:InitialConditionNonZeroB}
\nu^f(0)=\begin{pmatrix}
\zeta_1\\
\zeta_2\\
\zeta_3\\
\zeta_4
\end{pmatrix},
\qquad \mid \zeta_1\mid^2 +\mid \zeta_2\mid^2 +\mid \zeta_3\mid^2
+\mid \zeta_4\mid^2=1,
\end{equation}
the solution of Eq.~(\ref{eq:SystemNonZeroB}) can be presented in the form
\begin{equation}
\label{eq:SolutionGeneralNonZeroB}
\nu^f(t)=\begin{pmatrix}
C_{11} & C_{12} & C_{13} & C_{14}\\
C_{21} & C_{22} & C_{23} & C_{24}\\
C_{31} & C_{32} & C_{33} & C_{34}\\
C_{41} & C_{42} & C_{43} & C_{44}
\end{pmatrix}
\begin{pmatrix}
e^{-i\tilde\omega_1 t}\\
e^{-i\tilde\omega_2 t}\\
e^{-i\tilde\omega_3 t}\\
e^{-i\tilde\omega_4 t}
\end{pmatrix},
\end{equation}
where the integration constants $C_{nm}$ are given by the following expressions:
\begin{align}
\label{eq:ConstantsNonZeroB}
C_{n1}&= - \,\frac{\beta_{n}^{(4)}+\beta_{n}^{(3)}\tilde\omega_1+
\beta_{n}^{(2)}(\tilde\omega_2\tilde\omega_3+\tilde\omega_2\tilde\omega_4+\tilde\omega_3\tilde\omega_4)+
\beta_{n}^{(1)}\tilde\omega_2\tilde\omega_3\tilde\omega_4}
{(\tilde\omega_1 -\tilde\omega_4)(\tilde\omega_1-\tilde\omega_3)(\tilde\omega_1-\tilde\omega_2)},
\\
C_{n2}&=-\frac{\beta_{n}^{(4)} +\beta_{n}^{(3)}\tilde\omega_2
+\beta_{n}^{(2)}(\tilde\omega_1\tilde\omega_3 +\tilde\omega_1\tilde\omega_4+\tilde\omega_3\tilde\omega_4)
+\beta_{n}^{(1)}\tilde\omega_1\tilde\omega_3\tilde\omega_4}
{(\tilde\omega_2-\tilde\omega_4)(\tilde\omega_2-\tilde\omega_3)(\tilde\omega_2-\tilde\omega_1)},
\\
C_{n3}&= -\, \frac{\beta_{n}^{(4)}
+\beta_{n}^{(3)}\tilde\omega_3 +
\beta_{n}^{(2)}(\tilde\omega_1\tilde\omega_2+\tilde\omega_1\tilde\omega_4+\tilde\omega_2\tilde\omega_4)
+\beta_{n}^{(1)}\tilde\omega_1\tilde\omega_2\tilde\omega_4}
{(\tilde\omega_3-\tilde\omega_4)(\tilde\omega_3-\tilde\omega_2)(\tilde\omega_3-\tilde\omega_1)},
\\
C_{n4}&=-\frac{\beta_{n}^{(4)}+\beta_{n}^{(3)}\tilde\omega_4
+\beta_{n}^{(2)}(\tilde\omega_1\tilde\omega_2+\tilde\omega_1\tilde\omega_3+\tilde\omega_2\tilde\omega_3)
+\beta_{n}^{(1)}\tilde\omega_1\tilde\omega_2\tilde\omega_3}
{(\tilde\omega_4-\tilde\omega_3)(\tilde\omega_4-\tilde\omega_2)(\tilde\omega_4-\tilde\omega_1)},
\end{align}
with
\begin{align}
\label{eq:An}
\beta_{n}^{(1)}=\zeta_n, \qquad \beta_{n}^{(2)}=\sum_{k=1}^4 h^f_{nk}\zeta_{k}, \qquad \beta_{n}^{(3)}=\sum_{k,l=1}^4
h^f_{nk}h^f_{kl}\zeta_l, \qquad
\beta_{n}^{(4)}=\sum_{k,l,m=1}^4 h^f_{nk}h^f_{kl}h^f_{lm}\zeta_m.
\end{align}
Here $h^f_{nk}$ are the matrix elements of the effective Hamiltonian $H^f_{\rm eff}$.

%


%
%

\begin{thebibliography}{99}
%
\bibitem{Nobel2015} Y. Fukuda \emph{et al.} (Super-Kamiokande Collaboration), {Phys. Rev. Lett.} \textbf{81}, 1562 (1998); Q. R. Ahmad \emph{et al.} (SNO Collaboration), {Phys. Rev. Lett.} \textbf{87}, 071301 (2001); Q. R. Ahmad \emph{et al.} (SNO Collaboration),  Phys. Rev. Lett. \textbf{89}, 011301 (2002).
%
\bibitem{fujikawa80} K. Fujikawa and R. E. Shrock, {Phys. Rev. Lett.} \textbf{45}, 963 (1980).
%
\bibitem{bib:Giunti} C. Giunti and A. Studenkin, Rev. Mod. Phys. \textbf{87},
531 (2015).
%
\bibitem{andp16} C. Giunti, K. A. Kouzakov, Y.-F. Li, A. V. Lokhov, A. I. Studenikin, and S. Zhou, {Ann. Phys. (Berlin)} \textbf{528}, 198 (2016).
%
\bibitem{Cisneros:1970nq}
A.~Cisneros,
  Astrophys. Space Sci.  {\bf 10}, 87 (1971).
\bibitem{Schechter:1981hw}
  J.~Schechter and J.~W.~F.~Valle,
  Phys. Rev. D {\bf 24}, 1883 (1981).
\bibitem{Okun:1986hi}
  L. B. Okun, M. B. Voloshin, and M. T. Vysotsky, Yad. Fiz. {\bf 44}, 677 (1986) [Sov. J. Nucl. Phys. {\bf 44}, 440 (1986)].
\bibitem{Akhmedov:1988uk}
  E. Kh.~Akhmedov,
  Phys.\ Lett.\ B {\bf 213}, 64 (1988).
\bibitem{Lim:1987tk}
  C.~S.~Lim and W.~J.~Marciano,
  Phys.\ Rev.\ D {\bf 37}, 1368 (1988).
\bibitem{Akhmedov:1988hd}
  E. Kh.~Akhmedov and M. Yu.~Khlopov,
  Mod. Phys. Lett. A {\bf 3}, 451 (1988).
\bibitem{Vidal:1990fr}
  J.~Vidal and J.~Wudka,
  Phys. Lett. B {\bf 249}, 473 (1990).
\bibitem{Smirnov:1991ia}
  A.~Yu.~Smirnov,
  Phys. Lett. B {\bf 260}, 161 (1991).
\bibitem{Akhmedov:1993sh}
  E.~Kh.~Akhmedov, S.~T.~Petcov, and A.~Yu.~Smirnov,
  Phys. Rev. D {\bf 48}, 2167 (1993).
\bibitem{Likhachev:1990ki}
  G.~G.~Likhachev and A.~I.~Studenikin,
  Zh. Eksp. Teor. Fiz. {\bf 108}, 769 (1995) [J. Exp. Theor. Phys. {\bf 81}, 419 (1995)].
\bibitem{Dvornikov:2007aj}
  M.~Dvornikov,
  J. Phys. G {\bf 35}, 025003 (2008).
\bibitem{Dmitriev:2015ega}
  A.~Dmitriev, R.~Fabbricatore, and A.~Studenikin,
  PoS \textbf{CORFU2014}, 050 (2015).
\bibitem{Studenikin:2016zdx}
  A.~Studenikin,
  EPJ Web Conf. {\bf 125}, 04018 (2016).
\bibitem{Egorov:1999ah}
  A. M.~Egorov, A. E.~Lobanov, and A. I.~Studenikin,
Phys. Lett. B {\bf 491}, 137 (2000).
\bibitem{Lobanov:2001ar}
  A. E.~Lobanov and A. I.~Studenikin,
    Phys. Lett. B {\bf 515}, 94 (2001).
\bibitem{Dvornikov:2001ez}
  M.~S.~Dvornikov and A.~I.~Studenikin,
  Yad. Fiz. {\bf 64}, 1705 (2001) [Phys. At. Nucl. {\bf 64}, 1624 (2001)].
\bibitem{Dvornikov:2004en}
  M.~S.~Dvornikov and A.~I.~Studenikin,
  Yad. Fiz. {\bf 67}, 741 (2004) [Phys. At. Nucl. {\bf 67}, 719 (2004)].
%
\bibitem{GZK66} K. Greisen, {Phys. Rev. Lett.} \textbf{16}, 748 (1966); G. T. Zatsepin and V. A. Kuz'min, ZhETF Pis'ma {\bf 4}, 114 (1966) [{JETP Lett.} \textbf{4}, 78 (1966)].
%
\bibitem{Halzen_2016} F. Halzen, Nat. Phys. \textbf{13}, 232 (2017).
%
\bibitem{Raffelt99} G. G. Raffelt, Phys. Rep. \textbf{320}, 319 (1999).
%
\bibitem{Beck09} R. Beck, Astrophys. Space Sci. Trans. \textbf{5}, 43 (2009).
%
%
\bibitem{bib:Fabbricatore} R. Fabbricatore, A. Grigoriev, and A. Studenikin, J. Phys.: Conf. Ser. \textbf{718}, 062058 (2016).
%
\bibitem{bib:Wolfenstein} L. Wolfenstein, Phys. Rev. D \textbf{17}, 2369 (1978).
%
\bibitem{NASA_ISMF} E. J. Zirnstein, J. Heerikhuisen, H. O. Funsten, G. Livadiotis, D. J. McComas, and N. V. Pogorelov, Astrophys. J. Lett. \textbf{818}, L18 (2016).
%
\bibitem{Viaux_2013} N. Viaux, M. Catelan, P. B. Stetson, G. G. Raffelt, J. Redondo,
A. A. R.Valcarce, and A.Weiss, Astron. Astrophys. \textbf{558}, A12 (2013).
%
\bibitem{PDG2016} C. Patrignani \emph{et al.} (Particle Data Group), Chin. Phys. C \textbf{40}, 100001 (2016).
%
\bibitem{Meures_book} T. Meures, \emph{Development of a Sub-glacial Radio Telescope for the Detection
of GZK Neutrinos} (Springer International Publishing, Switzerland, 2015) pp.~13--23.
%
\bibitem{bib:Ishihara} A. Ishihara, J. Phys.: Conf. Ser. \textbf {718}, 062027 (2016).
%
\bibitem{Zadeh_book} L. A. Zadeh and C. A. Desoer, \emph{Linear System Theory: The State Approach} (McGraw Hill, New York, 1963) pp.~303--305.
%
\bibitem{Abramowitz_book} \emph{Handbook of Mathematical Functions with Formulas, Graphs, and Mathematical Tables, 9th printing}, edited by M.~Abramowitz and I.~A.~Stegun (Dover, New York, 1972) pp.~17--18.
%
\bibitem{Beyer_book} W.~H. Beyer, \emph{CRC Standard Mathematical Tables, 28th ed.} (CRC Press, Boca Raton, FL, 1987) p.~12.
%
\end{thebibliography}
\end{document}